\documentclass[8pt,a4paper,twocolumn]{article}
\usepackage[margin=20mm]{geometry}
\usepackage{cite}
\usepackage{amsmath,amssymb,amsfonts}
\usepackage{algorithmic}
\usepackage{graphicx}
\usepackage{textcomp}

\usepackage{mathtools, braket}
\usepackage[caption=false,font=footnotesize]{subfig}
\graphicspath{{./images/}}

\def\BibTeX{{\rm B\kern-.05em{\sc i\kern-.025em b}\kern-.08em
    T\kern-.1667em\lower.7ex\hbox{E}\kern-.125emX}}
\begin{document}

\title{Decomposition of Multi-Qubit Gates for Circuit Cutting}
\author{
    Ryota Tamura$^{1,\ast}$, Tomoya Kashimata$^{1}$, Yohei Hamakawa$^{1}$,\\
     Kosuke Tatsumura$^{1}$, and Hiroshi Imai$^{2}$\\
    \small$^{1}$Corporate Laboratory, Toshiba Corporation, 1 Komukai Toshiba-cho, Saiwai-ku, Kawasaki 212-8582, Japan\\
    \small$^{2}$Human Genome Center, The Institute of Medical Science, The University of Tokyo,\\
    \small 4-6-1 Shirokanedai, Minato-ku, Tokyo 108-8639, Japan\\
    \small${^{\ast}}$Corresponding author: Ryota Tamura (e-mail: ryota.tamura.d11@mail.toshiba)
}

\maketitle

\begin{abstract}
A large-scale quantum circuit can be partitioned into multiple subcircuits through circuit cutting, where each subcircuit is executed multiple times and the expectation value of the original circuit is reconstructed by classical post-processing from their measurement (sampling) results.
In this process, appropriate cut locations are identified after the user-designed quantum circuit, including multi-qubit gates that act on three or more qubits, has been decomposed into single-qubit gates and two-qubit gates such as the CNOT gate.
Here, we present a method for reducing the sampling overhead, which refers to the increase in the number of samples required due to the cutting process, by modifying the decomposition strategy of multi-qubit gates. Using MCX and CCCX gates as representatives of multi-qubit gates, we demonstrate that the proposed decomposition method, which introduces a small number of ancilla qubits according to the identified cut locations, effectively decreases the sampling overhead.
\end{abstract}

% main text
\section{Introduction}\label{sec:introduction}
While the number of qubits available in quantum computers continues to grow~\cite{IBMrm}, there remains a general need to solve problems that require more qubits than physical quantum devices can support. In this context, methods such as circuit cutting~\cite{Bravyi_2016,Peng_2020,Mitarai_2021,Mitarai_2021_2} have attracted significant attention.

Circuit cutting is a method that, given a quantum circuit and a classical function that computes an expectation value based on the circuit's measurement results, divides the original circuit into multiple subcircuits.
Each subcircuit is executed multiple times, and the expectation value corresponding to the original circuit is reconstructed through classical post-processing of the individual measurement (sampling) results.
For simplicity, this paper focuses on the case of bipartitioning.

\begin{figure}[t!]
  \centering
  % (a)
  \subfloat[Wire cutting where a quantum wire is cut in the middle of the circuit.]{
    \includegraphics[width=0.95\columnwidth]{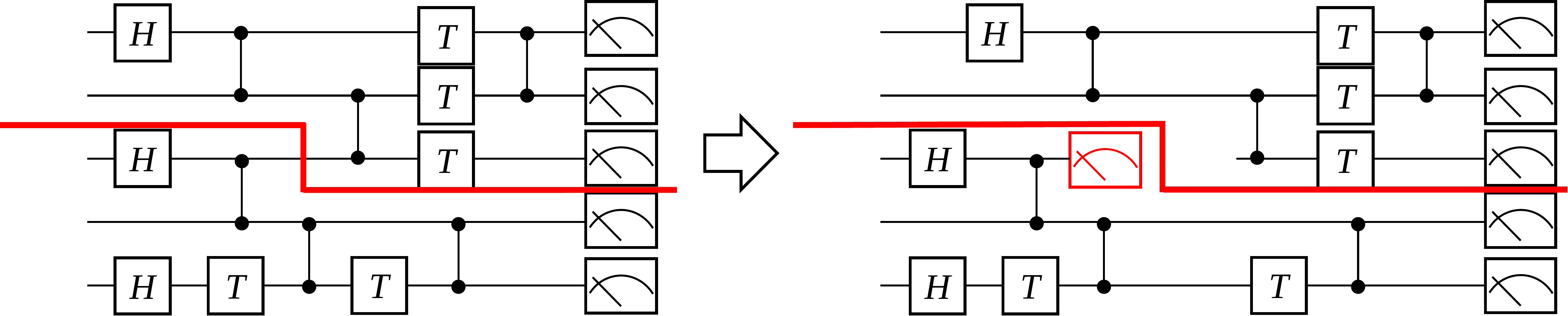}
  }\vspace{2pt}
  % (b)
  \subfloat[Gate cutting where a two-qubit or multi-qubit gate is cut.]{
    \includegraphics[width=0.95\columnwidth]{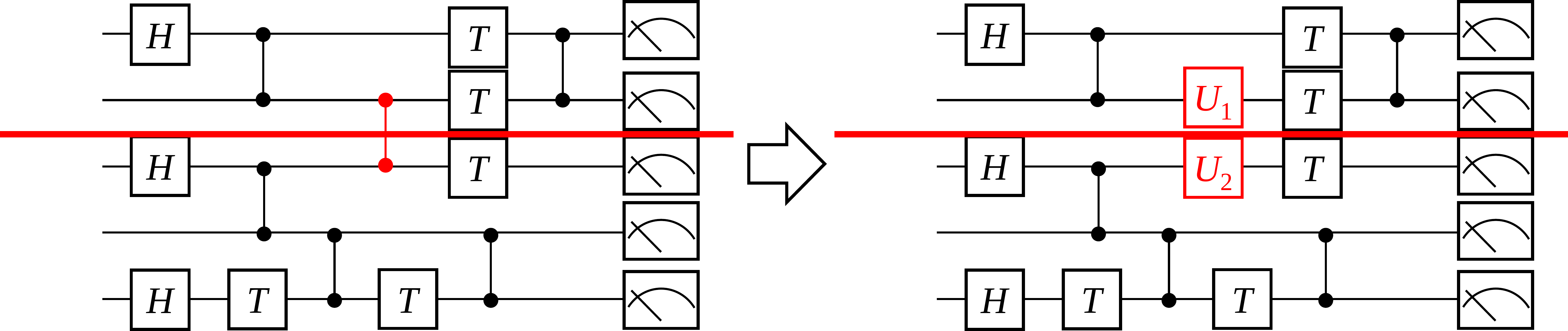}
  }
  \caption{Circuit cutting. A quantum circuit is partitioned into independently executable subcircuits along red line.\label{intro_1}}
\end{figure}

When partitioning a circuit, cutting a qubit wire in the middle of the circuit along the depth (time) direction is referred to as a wire cut (Fig.~\ref{intro_1}(a)).
On the other hand, partitioning a two-qubit (or multi-qubit) gate that acts on two (or more) qubits is referred to as a gate cut (Fig.~\ref{intro_1}(b)).
A circuit cutting method that enables wire cuts was proposed by Peng {\it et al.}~\cite{Peng_2020}.
Subsequently, a method that enables gate cuts was proposed by Mitarai {\it et al.}~\cite{Mitarai_2021, Mitarai_2021_2}.

\begin{figure*}[t]
\centering
\includegraphics[width=0.95\textwidth]{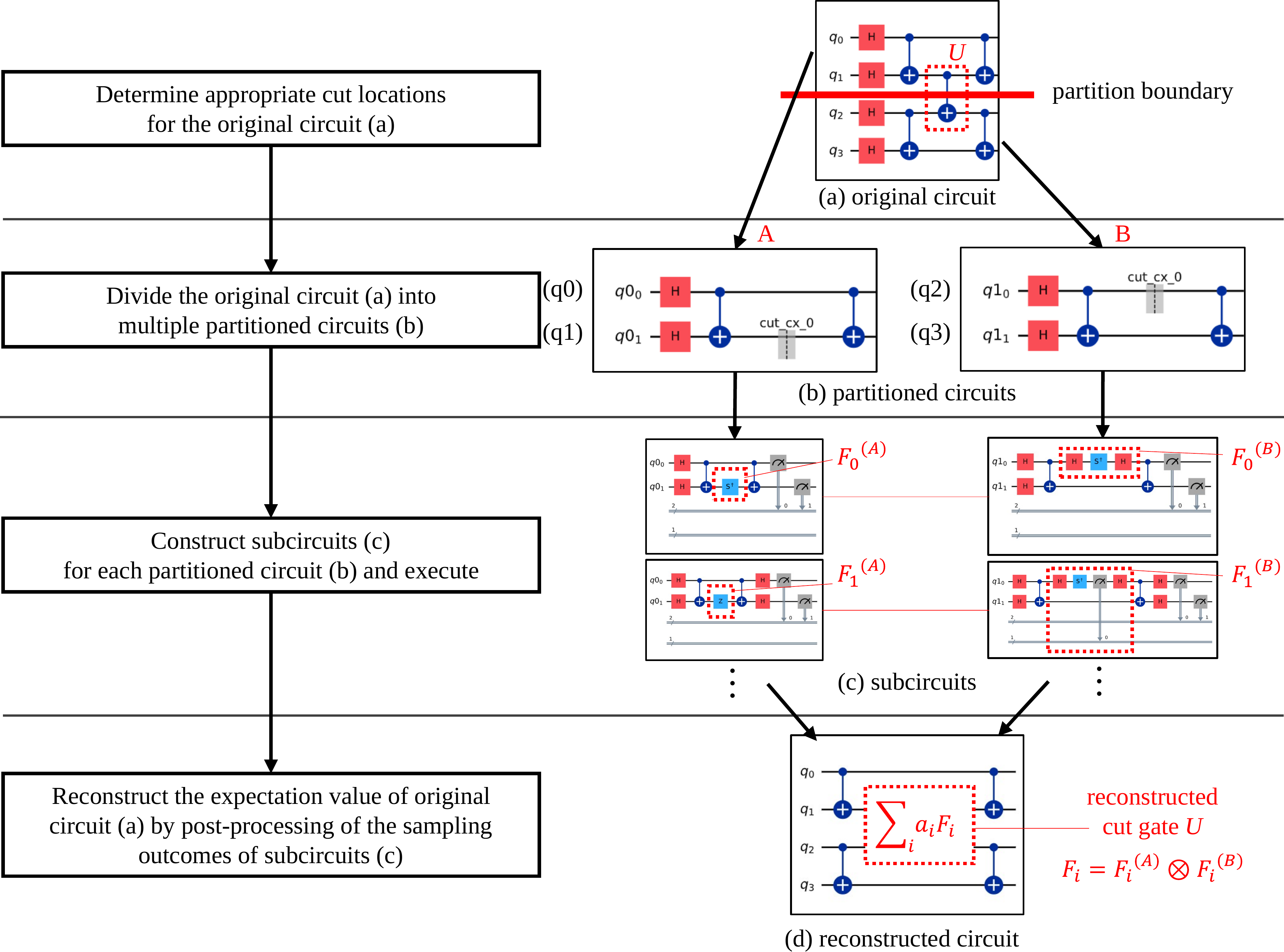}
\caption{Procedure for partitioning a quantum circuit and executing its subcircuits\label{intro_flow}}
\end{figure*}

Fig.~\ref{intro_flow} illustrates the procedure for partitioning a quantum circuit and executing the resulting subcircuits~\cite{qiskit-2025a, qiskit-2025b, Brandhofer_2024, Tang_2021}.
First, appropriate cut locations in the original circuit (Fig.~\ref{intro_flow}(a)) are determined.
In addition to manually specifying the assignment of qubits to each partitioned circuit, several methods have been proposed for automatically searching for suitable partitionings~\cite{qiskit-2025a, qiskit-2025b, Brandhofer_2024, Tang_2021}.
Second, the original circuit is divided into two partitioned circuits (Fig.~\ref{intro_flow}(b)), denoted A and B.
During this step, every two-qubit gate and wire that crosses the boundary of partitioned circuits (referred to as the partition boundary) is replaced with cut gates (e.g. cut\_cx\_0) and cut wires, which cannot be executed directly.
Third, for each cut gate $U$ that crosses the partition boundary, decomposition expressed by $U=\sum_ia_iF_i^{(A)} \otimes F_i^{(B)}$ is derived and the corresponding subcircuits (Fig.~\ref{intro_flow}(c)) on each side are constructed by replacing cut gates with ${F_i^{(A)}}$ in partition A and ${F_i^{(B)}}$ in partition B.
Let $I$ denote the index set labeling the terms in the decomposition of $U$.
This procedure yields subcircuits on sides A and B, which we organize into index-matched pairs ${(A_i,B_i)}_{i \in I}$.
Each index-matched pair is executed repeatedly to obtain sampling outcomes.
Finally, the expectation value of the original circuit is reconstructed (Fig.~\ref{intro_flow}(d)) by classical post-processing of the sampling outcomes from all index-matched pairs of subcircuits.

Post-processing in circuit cutting is based on quasi-probability simulation~\cite{Temme_2017,Endo_2018,Piveteau_2022}.
Consider a gate $U$ that crosses the partition boundary (Fig.~\ref{intro_flow}(a)).
Such a gate $U$ can be decomposed as $U=\sum_ia_iF_i$~\cite{Mitarai_2021,qiskit-2025a,Piveteau_2024,Ufrecht_2023}, where each $F_i$ is an operator that acts locally on the subcircuits (Fig.~\ref{intro_flow}(c)) and factorizes as $F_i=F_i^{(A)} \otimes F_i^{(B)}$.
Due to the linearity of quantum gates, the expectation value obtained when executing the original circuit can be reconstructed from the weighted sum over the outcomes of the index-matched pairs ${(A_i,B_i)}_{i \in I}$ with coefficients $a_i$.
Although the coefficients $a_i$ play a role analogous to probabilities, they may take negative values.
For this reason, the method is referred to as ``quasi''-probability simulation.

In circuit cutting, it is known that, to maintain a fixed level of accuracy in the reconstructed expectation value,
 the number of circuit executions (i.e., the number of samples) required increases exponentially with respect to the number of wire cuts and gate cuts.
More precisely, larger absolute values of the weights $a_i$ lead to a greater contribution to the final result.
Therefore, in order to keep the error below a fixed threshold, a number of samples proportional to $\sum_i |a_i|^2$, known as the sampling overhead factor, is required.
This value depends on the object being cut: gate cuts on a CNOT gate require $O(3^{2n})$ samples~\cite{Piveteau_2024},
 whereas wire cuts require $O(4^{2n})$ samples~\cite{brenner2023}, where $n$ denotes the number of cuts.
As the number of cuts increases, the computational speed advantage of quantum computers—which is the primary motivation for using quantum computers—deteriorates.
Therefore, methods for reducing the sampling overhead have been actively studied.

\begin{figure*}[t]
\centering
\includegraphics[width=0.95\textwidth]{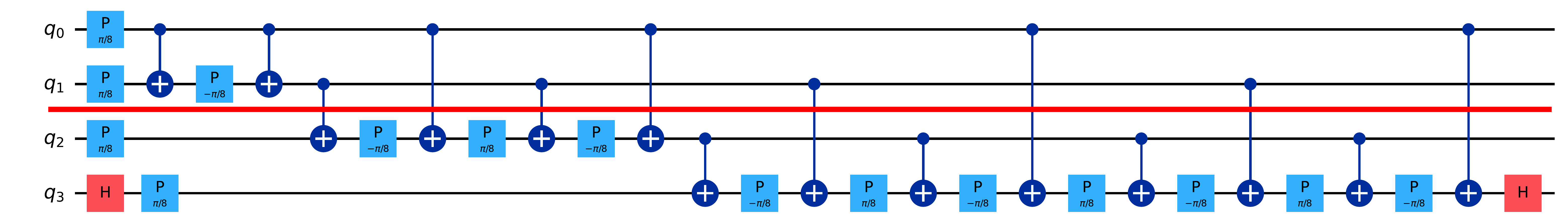}
\caption{Conventional decomposition of CCCX gate\label{CCCX_EGQCs7_1}}
\end{figure*}

% Prior Work
When classical communication between the partitioned circuits is allowed, a method has been proposed that reduces the sampling overhead using quantum teleportation~\cite{Piveteau_2024,brenner2023}.
In this method, the sampling overhead associated with gate cuts can be reduced from approximately $O(3^{2n})$ to $O(2^{2n})$ by treating multiple cuts collectively.

Several frameworks that, by optimizing the circuit partition, reduce the sampling overhead and improve the reliability of the obtained results have been proposed.
CutQC~\cite{Tang_2021} and ScaleQC~\cite{tang2022}, proposed by Tang {\it et al.}, are frameworks for circuit cutting, where the subcircuits executed on quantum hardware can become smaller, which reduces noise and consequently improves fidelity.
Ren {\it et al.} have proposed a framework that applies gate cuts while taking the hardware layout into account~\cite{Ren2024}.
By reducing the number of SWAP gates compared with CutQC, their method decreases the circuit depth and consequently improves the final reliability.
Ufrecht {\it et al.}\ have proposed a method for directly cutting MCZ (multi-controlled Z) gates~\cite{Ufrecht_2023}.
This method replaces the MCZ gate with a predefined sequence of post-cut gates ($F_i$).
In this method, the sampling overhead becomes $O(6^{2n})$ for an arbitrary number of control qubits,
 and it is reduced to $O(4.5^{2n})$ in the special case of a CCZ (Controlled-Controlled Z gate) gate, which has two control qubits.

%Introduction to Proposal
In current circuit cutting research, most studies have focused on gate cuts for CNOT and CZ gates, as well as wire cuts. Thus, these methods are applied only after the quantum circuit has been decomposed into single-qubit gates and two-qubit gates such as the CNOT gate.
Qiskit~\cite{qiskit} provides a circuit cutting module~\cite{qiskit-2025a,qiskit-2025b}, which applies cutting only after the circuit has been converted into basic gates using Qiskit's \verb|decompose()| method. Thus, it does not directly cut multi-qubit gates.

In this paper, we propose a method to handle MCX (multi-controlled X) gates (as a representative family of multi-qubit gates) for circuit cutting.
When the partition boundary crosses multi-qubit gates, the proposed method decomposes the multi-qubit gates by inserting ancilla qubits while taking into account the place of the partition boundary, and consequently reduces the sampling overhead of circuit cutting.
The proposed method can be applied in combination with existing circuit cutting techniques such as \cite{brenner2023, Piveteau_2024, Tang_2021, Brandhofer_2024}, because it operates between the identification of cut locations (Fig.~\ref{intro_flow}(a)) and the construction of partitioned circuits (Fig.~\ref{intro_flow}(b)).
As a result, it leaves the downstream cutting and post-processing frameworks unchanged and can be combined with them without modification.
As a representative example, we describe a decomposition method for the MCX gate and show that, combining with the method of Piveteau {\it et al.}~\cite{Piveteau_2024}, it reduces the sampling overhead compared with prior work~\cite{Ufrecht_2023}.
The MCX gate serves as a fundamental building block for implementing classical logic within quantum circuits~\cite{Wille2008}, and its efficient realization is crucial because it directly impacts the overall implementation cost of the circuit.
Before describing the MCX gate, we also describe the CCCX (controlled-controlled-controlled X) gate as a more specific example for better understanding.

\section{Proposed Method}\label{cpt:proposal}
\subsection{Decomposition of CCCX Gate}
When the MCX gate has three control qubits, it is called CCCX (controlled-controlled-controlled X) gate. We first explain the proposed decomposition method for the CCCX gate and then move on to more general MCX gates.

There are various decomposition methods for the CCCX gate. If circuit cutting is not taken into consideration,
 it is typically chosen so that the number of required two-qubit gates or the circuit depth are minimized.
An example of a decomposition that minimizes the number of two-qubit gates is shown in Fig.~\ref{CCCX_EGQCs7_1}\cite{Barenco_1995}.
In this decomposition, the CCCX gate can be implemented using 14 CNOT gates.
However, if the circuit is partitioned between the second and third qubits ($q_1$ and $q_2$), eight of these CNOT gates cross the partition boundary,
 leading to an overhead of $O(3^{16})$ under gate cuts.
This motivates a partition-aware decomposition that reduces the number of cuts.

\begin{figure}[t]
\centering
\includegraphics[width=0.95\columnwidth]{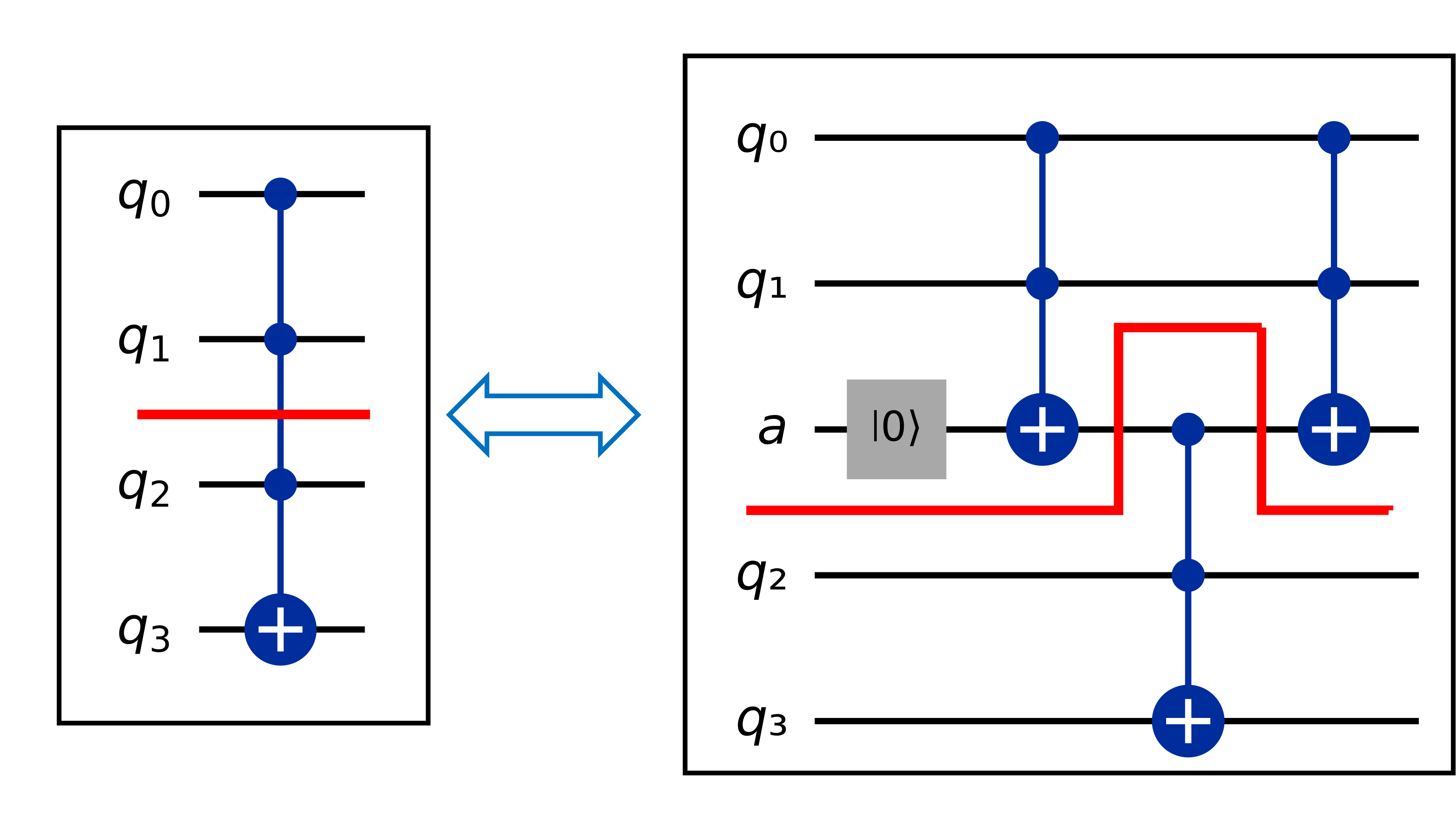}
\caption{Decomposition of CCCX gate using one ancilla qubit ($a$)\label{CCCX_dec1}}
\end{figure}

\begin{figure}[t]
\centering
\includegraphics[width=0.95\columnwidth]{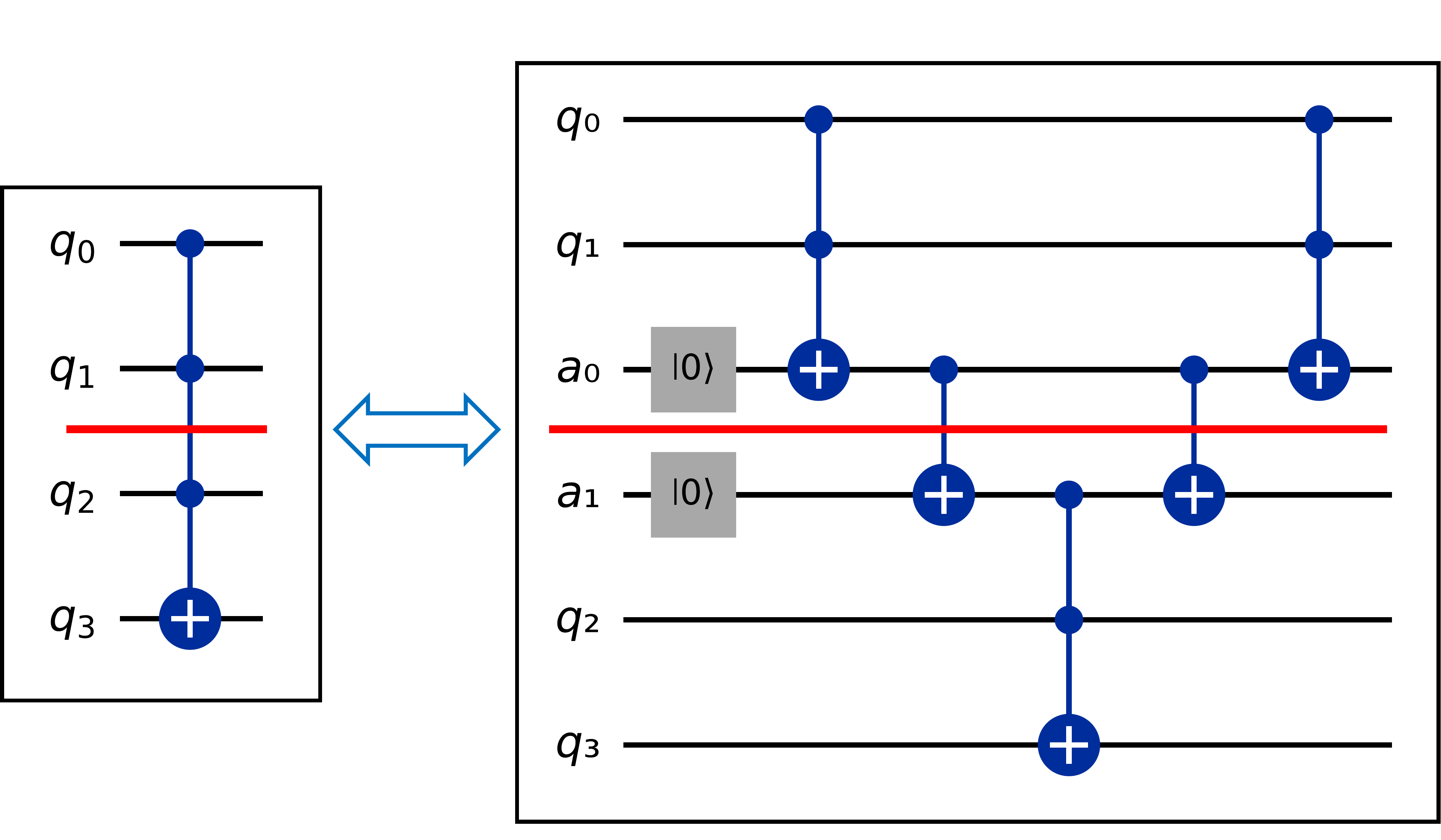}
\caption{Decomposition of CCCX gate using two ancilla qubits ($a_0$ and $a_1$) (dec2A)\label{CCCX_dec2A}}
\end{figure}

In the case where the circuit is partitioned between the second and third qubits of a CCCX gate,
 the gate can be equivalently transformed into the circuit shown in Fig.~\ref{CCCX_dec1}.
In this decomposition, an ancilla qubit $a$ initialized to $\ket{0}$ is added to the circuit, and the CCCX gate is decomposed into three CCX (Toffoli) gates, each having one fewer control qubit.
The first and third Toffoli gates share the same configuration: $q_0$ and $q_1$ are used as the two control qubits and $a$ is the target qubit.
After applying the first Toffoli gate, the ancilla qubit $a$ is in a dirty state, meaning that it contains intermediate computational information, then the third Toffoli gate uncomputes $a$, returning it to the clean $\ket{0}$ state for later reuse in the circuit.
The second Toffoli gate uses $a$ and $q_2$ as the two control qubits and $q_3$ as the target qubit.

The CCCX gate is also equivalent to the circuit shown in Fig.~\ref{CCCX_dec2A}.
In this decomposition, an ancilla qubit $a_0$ is added to the partition that contains only control qubits ($q_0$ and $q_1$) and another ancilla qubit $a_1$ is added to the partition that contains the target qubit $q_3$.
Two CNOT gates are then applied, using $a_0$ as the control qubit and $a_1$ as the target qubit; the second CNOT gate uncomputes $a_1$, returning it to the clean $\ket{0}$ state.
Because six CNOT gates are required~\cite{Barenco_1995} to implement a Toffoli gate, the number of CNOT gates needed to realize the CCCX gate is 18 in the circuit shown in Fig.~\ref{CCCX_dec1} and 20 in the circuit shown in Fig.~\ref{CCCX_dec2A}.
These numbers are larger than the 14 CNOT gates required in the decomposition of Fig.~\ref{CCCX_EGQCs7_1}, however this increase in the nuber of CNOT gates can be mitigated by using logic-circuit synthesizing methods.

Interestingly, the sampling overhead for these decomposition methods is largely different as follows. When the circuit is partitioned between the second and third qubits of the original CCCX gate,
 the circuit in Fig.~\ref{CCCX_dec1} incurs an overhead of $O(4^{4})$ because two wires are cut,
 whereas the circuit in Fig.~\ref{CCCX_dec2A} incurs an overhead of $O(3^{4})$ because two CNOT gates are cut.
Both values are significantly smaller than the $O(3^{14})$ overhead associated with the decomposition in Fig.~\ref{CCCX_EGQCs7_1}.

\begin{figure}[t]
\centering
\includegraphics[width=0.95\columnwidth]{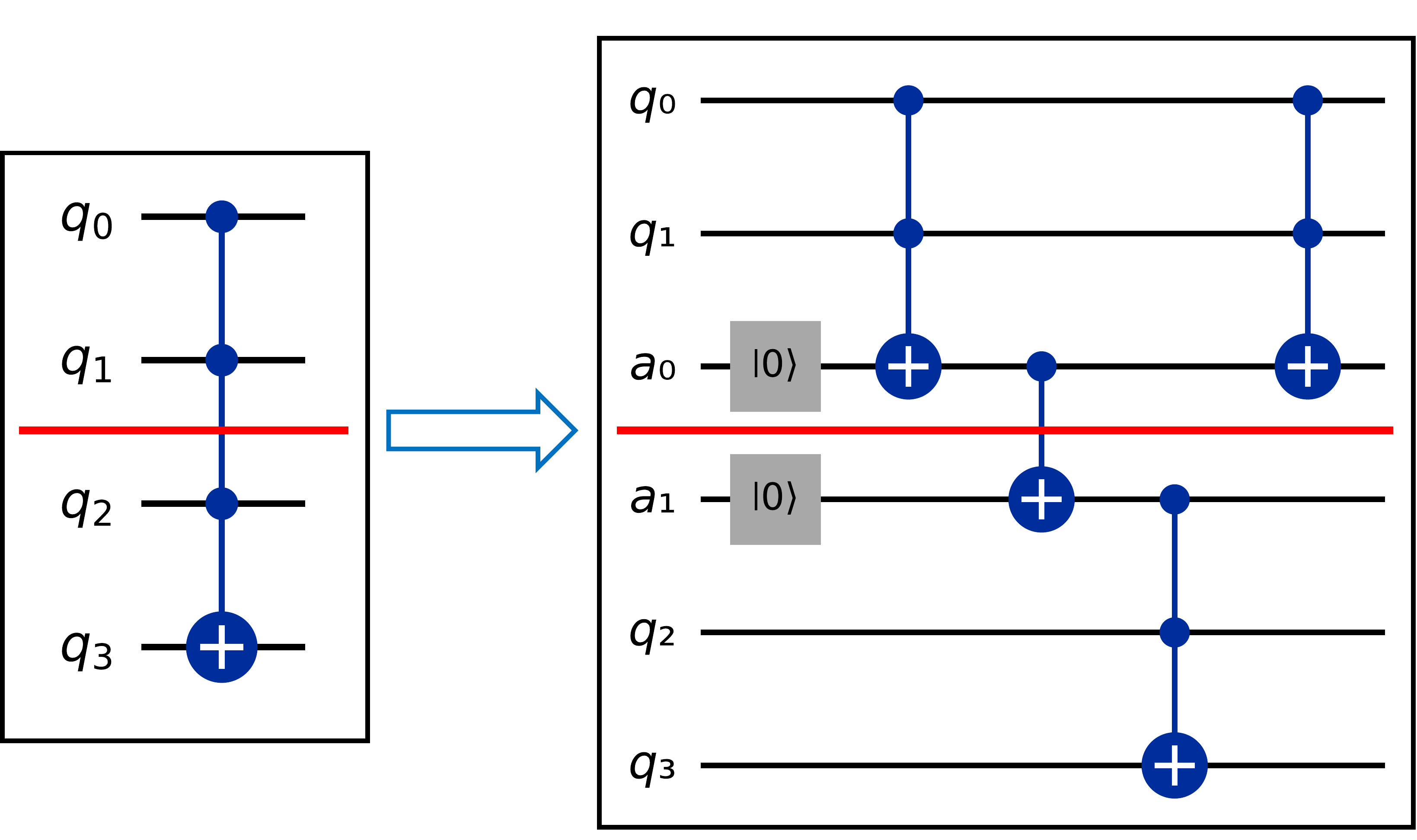}
\caption{Decomposition of CCCX gate using two ancilla qubits ($a_0$ and $a_1$), with one ancilla qubit ($a_1$) left in a dirty state (dec2Ad)\label{CCCX_dec2Ad}}
\end{figure}

By removing the second CNOT gate of the dec2A (Fig.~\ref{CCCX_dec2A}), an alternative decomposition shown in Fig.~\ref{CCCX_dec2Ad}, in which the number of cut CNOT gates is reduced to one, can be obtained.
Although the ancilla qubit $a_1$ in the partitioned circuit containing the target qubit remains in a dirty state and therefore cannot be reused, the sampling overhead is further reduced to $O(3^{2})$ compared with that of dec2A (Fig.~\ref{CCCX_dec2A}).
While this decomposition is not strictly equivalent to the original CCCX gate, the difference in action between CCCX gate and dec2Ad (Fig.~\ref{CCCX_dec2Ad}) is limited to the state of the ancilla qubit $a_1$, as discussed in detail in Section~\ref{cpt:proposal_2}.

\subsection{Decomposition of MCX Gate}\label{cpt:proposal_2}
The decomposition method for the CCCX gate described in the previous subsection is extended to MCX gates,
 and a mathematical verification of the decomposition is presented.

\begin{figure}[t]
\centering
\includegraphics[width=0.95\columnwidth]{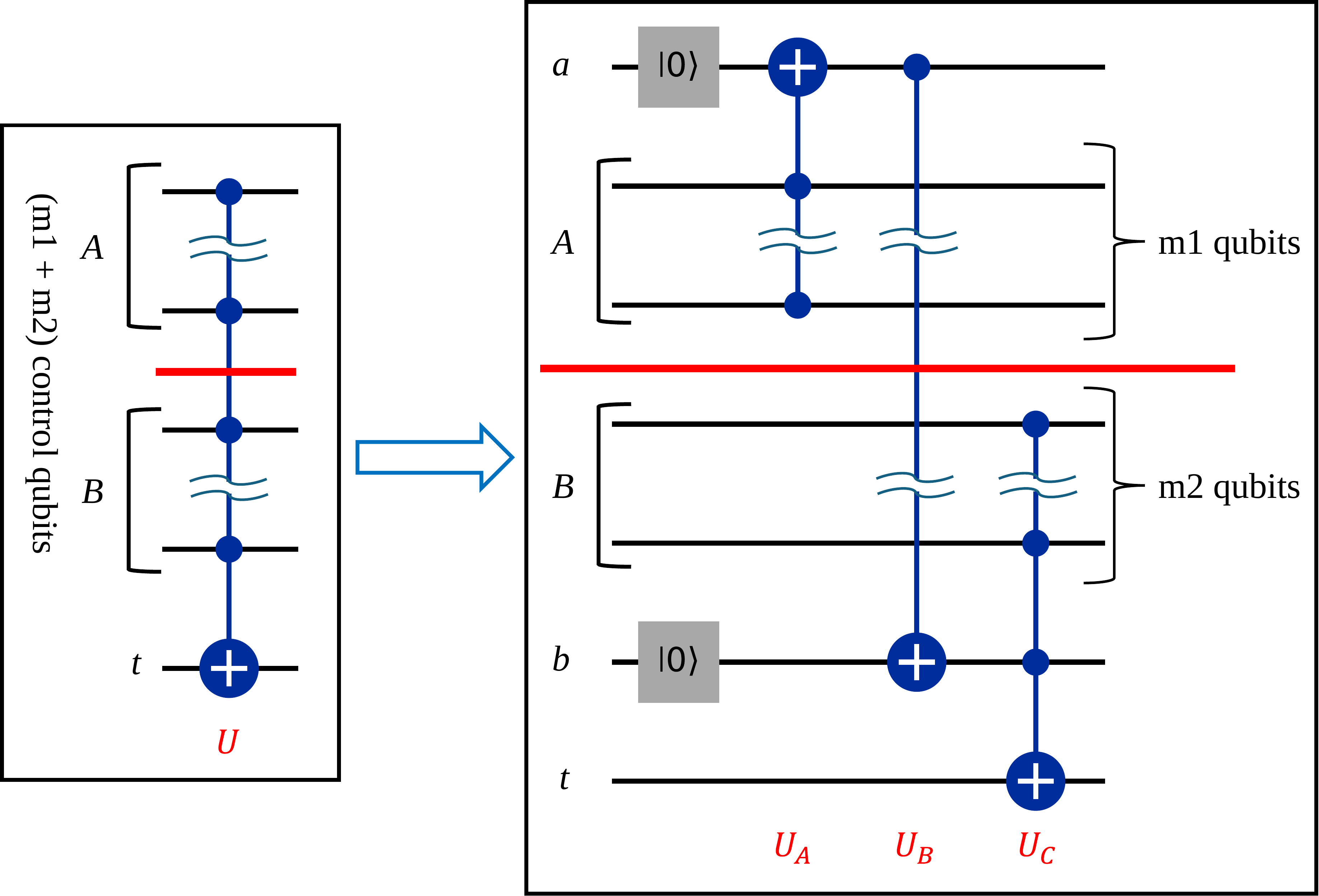}
\caption{Decomposition of MCX gate using two ancilla qubits ($a$ and $b$), with two ancilla qubits ($a$ and $b$) left in dirty states (dec2Ad). The original MCX gate ($U$) is replaced by two MCX gates ($U_A$ and $U_C$) and one CNOT gate ($U_B$).\label{mcx_transform2}}
\end{figure}

The basic principle of the decomposition for the CCCX gate remains the same for MCX gates as shown in Fig.~\ref{mcx_transform2}.
Among the two partitioned circuits produced by circuit cutting, let $A$ denote the set of control qubits of the original MCX gate in the partition that contains only control qubits, and add an ancilla qubit $a$ initialized to $\ket{0}$ to that partition.
In the other partition, which contains the target qubit of the original MCX gate, let $B$ denote the control qubits and $t$ the target qubit, and add an ancilla qubit $b$.
Define the following gates: the MCX gate $U_A$, which uses $A$ as the control qubits and $a$ as the target qubit; the CNOT gate $U_B$, which uses $a$ as the control qubit and $b$ as the target qubit; and the MCX gate $U_C$, which uses $b$ and $B$ as the controls and $t$ as the target qubit.
However, for the sake of simplifying the formal verification, the ancilla qubits are placed above and below the control qubits rather than directly on the partition boundary introduced by circuit cutting.

The action of the decomposed quantum circuit is expressed using matrix formulas based on tensor products,
 and its behavior is evaluated to verify the equivalence.
As shown in Fig.~\ref{mcx_transform2}, the entire quantum circuit consists of the ancilla qubits $a$ and $b$,
 the $m_1$ control qubits $A$, the $m_2$ control qubits $B$, and the target qubit $t$.
To represent the action of each gate, the projection operator $P$ is defined as follows,
 and the identity matrix $I$ and the Pauli matrix $X$ are used.

\begin{align}
P_A &:= (\ket{1}\!\bra{1})^{\otimes m_1}_{A}\label{eq:defPA},\\
P_{aA} &:= P_a \otimes P_A, \\
I_A &:= I^{\otimes m_1}_{A}, \\
I_{aA} &:= I_a \otimes I_A, \\
X_a &:= (\ket{0}\!\bra{1} + \ket{1}\!\bra{0})_{a}
\end{align}

\begin{table*}[tb]
\caption{Sampling overhead by circuit cutting}
\label{tab:evaluation}
\centering
\begin{tabular}{l|cc||cc|cc}
\hline
 & \multicolumn{2}{c||}{Number of qubits} 
 & \multicolumn{4}{c}{Sampling overhead} \\
 & & 
 & \multicolumn{2}{c|}{Without classical communication}
 & \multicolumn{2}{c}{With classical communication} \\
 & CCCX & MCX & CCCX & MCX & CCCX & MCX \\
\hline
Prior work\cite{Ufrecht_2023} 
    & 4 & $m$ 
    & $O(6^{2n})$ & $O(6^{2n})$
    & $O(6^{2n})$ & $O(6^{2n})$ \\
dec2A(Fig.~\ref{CCCX_dec2A}) 
    & 6 & $m+2$ 
    & $O(9^{2n})$ & $O(9^{2n})$
    & $O(4^{2n})$ & $O(4^{2n})$ \\
dec2Ad(Figs.~\ref{CCCX_dec2Ad} and \ref{mcx_transform2}) 
    & 6 & $m+2$
    & $O(3^{2n})$ & $O(3^{2n})$
    & $O(2^{2n})$ & $O(2^{2n})$ \\
\hline
\end{tabular}
\end{table*}

The subscripts indicate the qubits on which the operators act.
When multiple qubits appear in the subscript, as in $P_{aA}$, the notation represents
 the tensor product of the operators acting on each corresponding qubit.

Using the above expressions, the action of the decomposed sequence of gates, 
 $U' = U_C \cdot U_B \cdot U_A$, can be expressed as follows.

\begin{align}
U'=&\Big\lbrace I_{aA} \otimes \big\lbrack I_{Bbt} + P_{Bb} \otimes (X_t - I_{t}) \big\rbrack \Big\rbrace \notag\\
& \cdot \Big\lbrace \big\lbrack I_{aABb} + P_{a} \otimes I_{AB} \otimes (X_b - I_b) \big\rbrack \otimes I_{t} \Big\rbrace \notag\\
& \cdot \Big\lbrace \big\lbrack I_{aA} + (X_a - I_a) \otimes P_A\big\rbrack \otimes I_{Bbt} \Big\rbrace\\
=&\big\lbrack I_{aABbt} + I_{aA} \otimes P_{Bb} \otimes (X_t - I_{t}) \big\rbrack \notag\\
& \cdot \big\lbrack I_{aABbt} + P_{a} \otimes I_{AB} \otimes (X_b - I_b) \otimes I_{t} \big\rbrack \notag\\
& \cdot \big\lbrack I_{aABbt} + (X_a - I_a) \otimes P_A \otimes I_{Bbt} \big\rbrack\\
=&I_{aABbt} \notag\\
&  + (X_a - I_a) \otimes P_A \otimes I_{Bbt} \notag\\
&  + P_{a} \otimes I_{AB} \otimes (X_b - I_b) \otimes I_{t} \notag\\
&  + P_{a} (X_a - I_a)  \otimes P_A \otimes I_{B} \otimes (X_b - I_b) \otimes I_{t} \notag\\
&  + I_{aA} \otimes P_{Bb} \otimes (X_t - I_{t}) \notag\\
&  + (X_a - I_a) \otimes P_A \otimes P_{Bb} \otimes (X_t - I_{t}) \notag\\
&  + P_{a} \otimes I_{A} \otimes P_{B} \otimes P_{b} (X_b - I_b)  \otimes (X_t - I_{t}) \notag\\
&  + P_{a} (X_a - I_a) \otimes P_A \otimes P_{B} \otimes P_{b}  (X_b - I_b) \notag\\
&    \otimes (X_t - I_{t})\label{eq:decU}
\end{align}

Let the initial state of the circuit be
 $\ket{\psi} = \ket{0_a} \otimes \ket{\phi_A} \otimes \ket{\phi_B} \otimes \ket{0_b} \otimes \ket{\phi_t}$.
Then, the state after executing the decomposed circuit, $U' \ket{\psi}$, is computed as follows.
From the definition of $P$, the third, fifth, sixth, and seventh terms of \eqref{eq:decU},
 in which $P_a$ and $P_b$ appear individually with respect to the ancilla qubits $a$ and $b$, vanish because $\braket{1|0} = 0$.

\begin{align}
U' \ket{\psi} =&\big\lbrack I_{aABbt}  \notag\\
&  + (X_a - I_a) \otimes P_A \otimes I_{Bbt}  \notag\\
&  + P_{a} (X_a - I_a)  \otimes P_A \otimes I_{B} \otimes (X_b - I_b) \notag\otimes I_{t} \\
&  + P_{a} (X_a - I_a) \otimes P_A \otimes P_{B} \otimes P_{b}  (X_b - I_b)  \notag\\
&    \otimes (X_t - I_{t}) \big\rbrack \cdot \ket{\psi} \\
=& \ket{0_a} \otimes \ket{\phi_A} \otimes \ket{\phi_B} \otimes \ket{0_b} \otimes \ket{\phi_t}  \notag\\
&  + (\ket{1_a} - \ket{0_a}) \otimes P_A \ket{\phi_A} \otimes \ket{\phi_B} \otimes \ket{0_b} \otimes \ket{\phi_t}  \notag\\
&  + \ket{1_a} \otimes P_A \ket{\phi_A} \otimes \ket{\phi_B}  \otimes (\ket{1_b} - \ket{0_b}) \otimes \ket{\phi_t} \notag\\
&  + \ket{1_a} \otimes P_A \ket{\phi_A} \otimes P_{B} \ket{\phi_B} \otimes \ket{1_b}  \notag\\
&    \otimes (X_t - I_{t}) \ket{\phi_t} \\
=& \ket{0_a} \otimes     \ket{\phi_A} \otimes       \ket{\phi_B} \otimes \ket{0_b} \otimes \ket{\phi_t}  \notag\\
&  - \ket{0_a} \otimes P_A \ket{\phi_A} \otimes       \ket{\phi_B} \otimes \ket{0_b} \otimes \ket{\phi_t}  \notag\\
&  + \ket{1_a} \otimes P_A \ket{\phi_A} \otimes       \ket{\phi_B} \otimes \ket{1_b} \otimes \ket{\phi_t} \notag\\
&  + \ket{1_a} \otimes P_A \ket{\phi_A} \otimes P_{B} \ket{\phi_B} \otimes \ket{1_b}  \notag\\
&    \otimes (X_t - I_{t}) \ket{\phi_t}\label{eq:decUfin}
\end{align}

The first term of \eqref{eq:decUfin} indicates that the base quantum state remains unchanged from the input.
The second and third terms indicate that when $\ket{\phi_A}$ equals $\ket{1}^{\otimes m_1}$ relative to the base state,
 the ancilla qubits $a$ and $b$ take the value $\ket{1}$ instead of $\ket{0}$.
The final term indicates that when $\ket{\phi_A}$ equals $\ket{1}^{\otimes m_1}$ and $\ket{\phi_B}$ equals $\ket{1}^{\otimes m_2}$,
 the two ancilla qubits take the state $\ket{1}$, and the target qubit is flipped.
Therefore, it has been shown that the action of the decomposed gates $U'$ in this subsection agrees
 with that of the original MCX gate $U$, except for the states of the ancilla qubits.

\section{Performance Evaluation}
In this section, the sampling overhead of the proposed method and that of prior work are evaluated.

In the proposed method, after decomposing the multi-qubit gates,
 existing methods such as that of Piveteau {\it et al.}~\cite{Piveteau_2024} can be applied in combination in the stage where the gates and wires are cut.
Therefore, the sampling overhead per CNOT gate is taken to be $O(3^{2})$ when classical communication is not available and $O(2^{2})$ when it is available,
 and the overall sampling overhead is computed based on the number of gates that are cut under these two conditions.
By contrast, in the method of Ufrecht {\it et al.}~\cite{Ufrecht_2023}, an MCZ gate is directly replaced with a set of gates, therefore no gate cut can be formed after applying this method.
Consequently, the technique of Piveteau {\it et al.}~\cite{Piveteau_2024}—which reduces the sampling overhead by jointly handling cut CNOT gates—cannot be applied.
For both the proposed method and the prior work, the sampling overhead is computed for two types of gates—CCCX and MCX—by considering the case in which $n$ such gates are bipartitioned  into two subcircuits, each containing at least two qubits.

The results are summarized in Table~\ref{tab:evaluation}.
Although the method of Ufrecht {\it et al.}~\cite{Ufrecht_2023} targets MCZ gates,
 MCZ gates and MCX gates can be converted into each other by single-qubit gates,
 and therefore they are compared on an equal footing.
Methods that aim to minimize the number of CNOT gates (Fig.~\ref{CCCX_EGQCs7_1}) in the decomposition are not included,
 because the number of CNOT gates that are cut varies depending on the number of control qubits.

When the sampling overhead of the proposed method is compared with that of prior work,
 it is first observed that the dec2A (Fig.~\ref{CCCX_dec2A}) achieves an overhead of $O(4^{2n})$ in the case where classical communication is available.
This result indicates that, in scenarios where multiple gates or wires are cut rather than a single MCX gate alone, the proposed method outperforms the prior work.

In the case of the dec2Ad (Fig.~\ref{CCCX_dec2Ad}),
 it is observed that the sampling overhead achieves $O(3^{2n})$ even when a single MCX gate is cut or when classical communication is not used.
This indicates that the proposed method outperforms the prior work.
Furthermore, in the case where classical communication is available,
 the sampling overhead is reduced to $O(2^{2n})$, enabling even more efficient circuit cutting.

The proposed method reduces the sampling overhead by transforming the circuit, prior to applying circuit cutting,
 into a form in which the number of gates crossing the partition boundary is minimized.
In the case where an MCX gate is decomposed into a circuit consisting only of control qubits and another circuit that includes both the control and target qubits,
 the amount of control information that must be shared between the two circuits is no more than that represented by a single CNOT gate.
Moreover, because completely eliminating the information shared between the partitioned circuits (i.e., reducing it to zero) is impossible,
 the proposed method can be regarded as effectively minimizing the number of gates that cross the partition boundary.

\section{Conclusion and discussion}
We presented a method for reducing the sampling overhead induced by circuit cutting of large quantum circuits.
Our approach decomposes multi-qubit gates, with a focus on MCX gates, by introducing a small number of ancilla qubits and explicitly accounting for the context at the partition boundary.
The proposed partition-aware decomposition methods, dec2A and dec2Ad, reduce the sampling overhead for MCX gates from $O(6^{2n})$ to $O(4^{2n})$ and $O(2^{2n})$, respectively, when classical communication between the two subcircuits is permitted. Furthermore, we mathematically verified the equivalence between the original MCX gate and its decomposed implementation.

There are three possible topics for future research. The first issue is the development of a workflow that determines the circuit partition for circuit cutting and informs the multi-qubit gate decomposition. In conventional workflows such as Qiskit, circuit partition is typically performed after standard gate decomposition, which may result in cuts being placed at suboptimal locations. In addition, the suitability of the resulting decomposition and partition for mapping onto physical qubits remains unclear.
The second issue is to identify additional multi-qubit gates that essentially differ from MCX gates and frequently appear in quantum algorithms and then to investigate concrete decomposition methods for such gates.
The third issue is to further optimize the circuit obtained after applying the proposed decompositions.
Although these decompositions generally increase the number of CNOT gates, this increase may be mitigated by employing techniques such as the use of RTOF (Relative-phase Toffoli) gates~\cite{Clarino_2023} in subcircuits where relative phase changes are irrelevant.

\section{Acknowledgment}
This work was conducted under the UTokyo Quantum Initiative.

\section*{Competing interests}
R.T., T.K., Y.H., and K.T. are included in inventors on a Japanese patent application related to this work filed by the Toshiba Corporation (Japanese Patent Application No. 2025-150029). The authors declare that they have no other competing interests.

% Citations

\clearpage

\begin{figure}[!t]
\noindent\includegraphics[width=1in,height=1.25in,clip,keepaspectratio]{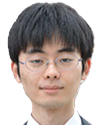}\\
{\small Ryota Tamura received the B.E. and M.E. degrees in computer science and engineering from Waseda University, Japan, in 2017 and 2019, respectively. He joined the Corporate Research and Development Center (currently, the Corporate Laboratory), Toshiba Corporation, Japan, in 2019. His research interests include Edge Cloud continuum, computer architecture and transpiler for quantum circuit.}
\end{figure}

\begin{figure}[!t]
\noindent\includegraphics[width=1in,height=1.25in,clip,keepaspectratio]{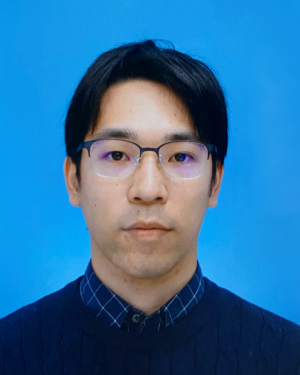}\\
{\small Tomoya Kashimata received the B.E. and M.E. degrees in computer science and engineering from Waseda University, Japan, in 2018 and 2020, respectively. He joined the Corporate Research and Development Center (currently, the Corporate Laboratory), Toshiba Corporation, Japan, in 2020. His research interests include Ising machine applications, computer architecture, and Processing-in-Memory.}
\end{figure}

\begin{figure}[!t]
\noindent\includegraphics[width=1in,height=1.25in,clip,keepaspectratio]{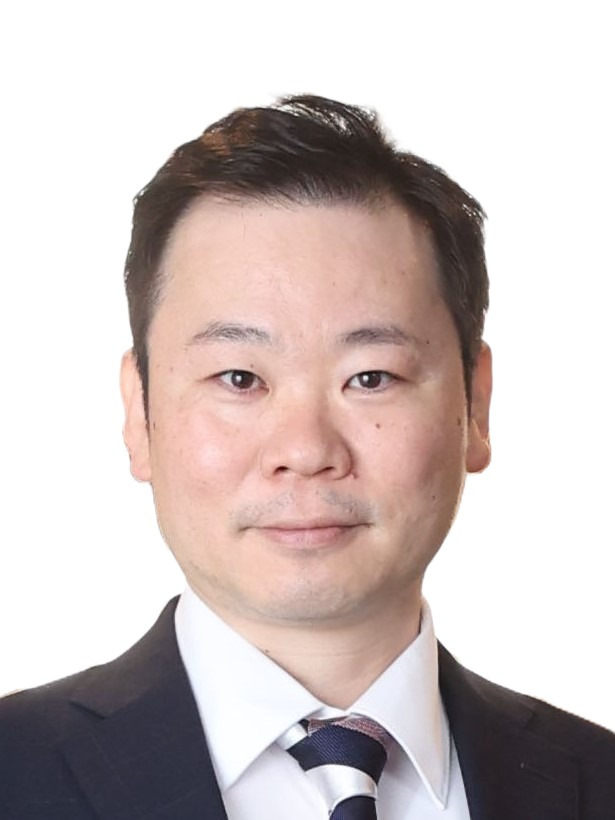}\\
{\small Yohei Hamakawa received the B.E and M.E. degrees in Advanced Electronics and Optical Science from Osaka University, Japan, in 2000 and 2002, respectively. He joined Toshiba Corporation in 2002. He was engaged in the development of image processing engines for digital televisions and TV products, and distributed computing algorithm in deep learning. His research interests include domain-specific computing, quantum computation, optimization in quantum circuit design, and their applications.}
\end{figure}

\begin{figure}[!t]
\noindent\includegraphics[width=1in,height=1.25in,clip,keepaspectratio]{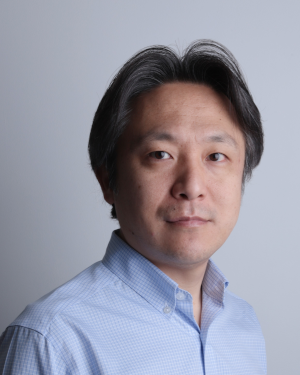}\\
{\small Kosuke Tatsumura received the B.E., M.E., and Ph.D. degrees in electronics, information and communications engineering from Waseda University, Japan, in 2000, 2001, and 2004, respectively. After working as a Postdoctoral Fellow with Waseda University, he joined the Corporate Research and Development Center (currently, the Corporate Laboratory), Toshiba Corporation, in 2006. There, he is currently a Senior Fellow, leading a research team and several projects toward realizing innovative industrial systems based on cutting-edge computing technology. He has been a Lecturer with Waseda University, since 2013. He was a Visiting Researcher with the University of Toronto, from 2015 to 2016. He received the Best Paper Award at IEEE International Conference on Field-Programmable Technology (FTP), in 2016, and the Awards for Science and Technology, the Commendation for Science and Technology by the Minister of Education, Culture, Sports, Science and Technology, Japan, in 2026.}
\end{figure}

\begin{figure}[!t]
\noindent\includegraphics[width=1in,height=1.25in,clip,keepaspectratio]{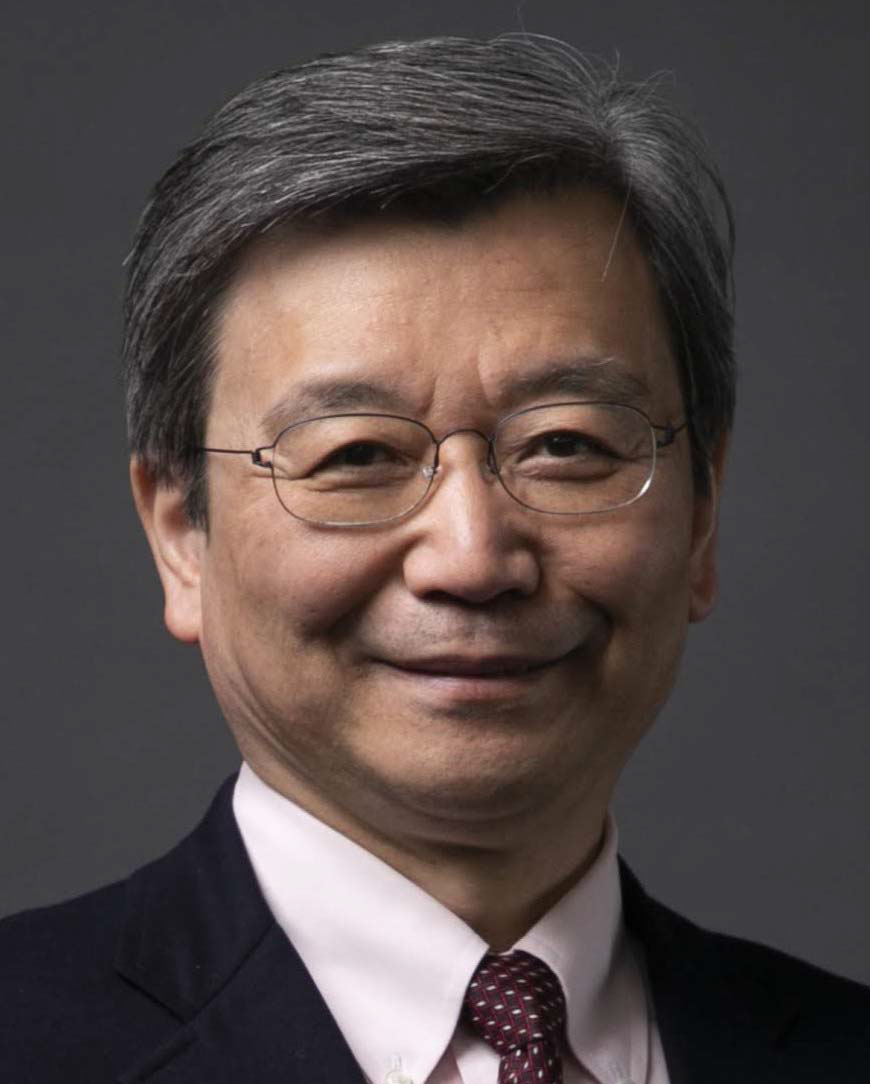}\\
{\small Hiroshi Imai received the B.Eng. degree in mathematical engineering and the D.Eng. degree in information engineering in 1981 and 1986, respectively. He became an Associate Professor at the Department of Information Engineering, Kyushu University, in 1986, and moved to the Department of Information Science, the University of Tokyo, in 1990, where he is currently a Professor with the Department of Computer Science. He led JSTERATO Quantum Computation and Information Project from 2000 to 2011. He is a leader of the field of quantum computing.}
\end{figure}

\end{document}